\documentclass[useAMS,usenatbib]{mn2e}
\usepackage{graphicx}
\usepackage{txfonts}

\title[Recurrent nova evolution from radio synchrotron emission]
{Insights into the evolution of symbiotic recurrent novae from radio synchrotron emission: V745 Scorpii and RS Ophiuchi}
\author[N. G. Kantharia et. al.]{N. G. Kantharia $^{1}$\thanks{E-mail:
\texttt{ngk@ncra.tifr.res.in}}, Prasun Dutta $^2$, Nirupam Roy$^3$, G. C. Anupama $^4$,  
\newauthor C. H. Ishwara-Chandra$^1$, A. Chitale, T. P. Prabhu $^4$\thanks{E-mail:\texttt{tpp@iiap.res.in}}, 
D. P. K.Banerjee$^5$, N. M. Ashok$^5$ \\
$^{1}$ National Centre for Radio Astrophysics, TIFR, Pune, India\\ 
$^{2}$ Indian Institute of Science Education and Research, Bhopal, India \\
$^3$ Indian Institute of Technology, Kharagpur, India \\
$^4$ Indian Institute of Astrophysics, Bangalore, India \\
$^5$ Physical Research Laboratory, Ahmedabad, India \\ }

\begin{document}

\date{Accepted 2015 October 7. Received 2015 August 19; in original form 2015 May 23}

\pagerange{\pageref{firstpage}--\pageref{lastpage}} \pubyear{2014}

\maketitle

\label{firstpage}

\begin{abstract}

We present  observations at 610 MHz and 235 MHz using the Giant Metrewave Radio Telescope (GMRT) of 
the recurrent nova V745 Scorpii which recorded its last outburst on 6 February 2014.  This is the second 
symbiotic recurrent nova whose light curve at low frequencies has been followed in detail, the 
first being RS Ophiuchi in 2006.  We fitted the 610 MHz light curve by a model of synchrotron emission 
from an expanding shell being modified by radiative transfer effects due to local absorbing gas consisting 
of a uniformly distributed and a clumpy component.  Using our model parameters, we find that the emission 
at 235 MHz peaked around day 35 which is consistent with our GMRT observations.  The two main results of 
our study are: (1) The radio emission at a given frequency is visible sooner after the outburst in 
successive outbursts of both V745 Scorpii and RS Ophiuchi.  The earlier detection of radio emission 
is interpreted to be caused by decreasing foreground densities. (2) The clumpy material, if exists, is close 
to the white dwarf and can be interpreted as being due to the material from the hot accretion disk.  The 
uniform density gas is widespread and attributed to the winds blown by the white dwarf.  We present 
implications of these results on the evolution of both novae. Such studies alongwith theoretical 
understanding have the potential of resolving several outstanding issues such as why all recurrent novae 
are not detectable in synchrotron radio and whether recurrent novae are progenitor systems of type 1a 
supernova.

\end{abstract}

\begin{keywords}
binaries: symbiotic -- novae -- radio continuum:stars  
\end{keywords}

\section{Introduction} Recurrent novae are binary systems with the primary being a white dwarf and the 
secondary star being a red giant or a main sequence star.   A symbiotic recurrent nova is a system 
comprising a red giant secondary.  The white dwarf accretes matter from the secondary which leads to a 
runaway thermonuclear explosion on its surface after a critical mass is reached.  This results in a 
spectacular increase in optical light by a few to 15 magnitudes. 

The last outburst of the recurrent nova V745 Sco, predicted by \citet{2010ApJS..187..275S} to be in year 
$2013\pm 1$, was recorded on 6/Feb/2014 (Rod Stubbings, AAVSO special notice 380).  The previous outbursts 
in V745 Sco were recorded on 10 May 1937 and 30 July 1989 \citep{1989Msngr..58...34D}. The current 
outburst has been detected in bands ranging from the $\gamma$ rays \citep{2014ATel.5879....1C}. to low 
frequency radio waves \citep{2014ATel.5962....1K}.
 
In this paper, we describe the low radio frequency observations of V745 Sco with GMRT and combine it with 
VLA archival data from the 1989 outburst to understand the evolution of the system. We also study the 
results on the symbiotic recurrent nova RS Ophiuchi from its outbursts in 1985 and 2006.

\section{Observations, data analysis and Results} The GMRT \citep{1991CuSc...60...95S} observations of 
V745 Sco were conducted either in the 610 MHz band or in the dual band which allows simultaneous 
observations in both the 235 MHz and 610 MHz bands between 7 February and 7 September 2014.  Bandwidth 
settings of 32 MHz at 610 MHz and 16 MHz at 235 MHz were used. A combination of Director's Discretionary 
Time and regular TAC-approved time in Observing Cycle 26 as part of the project {\it Galactic Novae with 
GMRT} (GNovaG) were used for GMRT observations.  The observations started with a 10 minute run on an 
amplitude calibrator (3C286 or 3C48) followed by 3 minute and 30 minute runs on the phase 
calibrator (J1830--360) and V745 Sco. Most of the observing slots were $\sim 2$h in duration  
with $\sim 1.5$h on the nova and $\sim 90~\mu$Jy rms noise on the final image at 610 MHz.

\begin{table*}
 \centering
  \caption{GMRT results on V745 Sco at 610 and 235 MHz.  $t_0$ is 6 February 2014.  The flux density
of the background source (bcksrc) is estimated from the image whereas the flux density of 
the phase calibrator J1830-360 is obtained from GETJY.}
  \begin{tabular}{@{}llrrrrrlrlrl@{}}
  \hline
   t--t$_0$     &  Date          & \multicolumn{4}{c}{Flux density at 610 MHz} 
    & \multicolumn{4}{c}{Flux density at 235 MHz } & $\alpha^{235}_{610}$ \\
  \hline
    &  &  V745Sco  & $\sigma$  & bcksrc  & J1830-360  &   V745Sco & $\sigma$ & bcksrc & J1830-360 &   \\
  days  & 2014 &  mJy/beam & mJy/beam &  mJy &  Jy &   mJy/beam & mJy/beam &  mJy & Jy &   \\
 \hline
3 & 9 Feb  & $<0.55$ & -  & 16.26(0.16) & 18.1(0.17)  & - & - & - & -  &\\
12 & 18 Feb & 1.07  & 0.14 & 16.41(0.14) & 16.87(0.05)  & - & - & - & - & \\
27 & 05 Mar & 6.9 & 0.13  &  17.04(0.13) & 18.07(0.11)  & - & - & - & -  &\\
28 & 06 Mar & 6.8  & 0.17   & 15.3(0.11) & 16.63(0.11)  & 4.3 & 0.9 & 30.9(0.9) & 29.44(0.3) & 0.49 \\
35 & 13 Mar & 5.89 & 0.18 & 14.76(0.1) & 16.27(0.16)  & 4.7 & 1.1 & 31.2(1.2) & 30.1(0.4) & 0.24 \\
40 & 18 Mar & 5.11 & 0.13 & 15.91(0.13) & 17.39(0.22)  & $<4.5$ & - & 28.7(1.6) & 32.5(0.7)  & $>0.13$ \\
47 & 25 Mar  & 3.83 & 0.18 & 15.64(0.18) & 16.95(0.15) & 5.4 & 1.4 & 30.7(1.4) & 31.9(0.7) & $-0.37$ \\
56 & 03 Apr  & 3.38 & 0.07 & 18.06(0.19) & 17.25(0.42) & $< 6$ & - & 43.1(0.98) & 30.73(2.11) & $>-0.6$ \\
71 & 18 Apr & 2.52  & 0.19 & 18.08(0.19) & 19.2(0.4) &  & - & - & - & \\
86 & 03 May & 2.06 & 0.07 & 17.59(0.07) & 18.54(0.08) & - & - & - & -  & \\
101 & 18 May  & 1.17  & 0.21 & 13.42(0.22) & 16.44(0.94) & $<4.5$ & & 25.1(1.15) & 28.44(0.29) & $>-1.4$ \\
123 & 09 June & 1.11 & 0.14  & 16.1(0.12) & 17.58(0.07) & - & - & - & & \\
154 & 10 July & 0.99 & 0.099 & 17.5(0.11) & 18.33(0.09) & - & - & - & & \\
187 & 12 Aug & 0.72  & 0.1  & 17.3(0.1) & 17.86(0.16)  & - & - & - & & \\
217 & 07 Sep & 0.48  & 0.12  & 16.6(0.07)  & 18.17(0.23)  & $<5.7$ & - & 26.3(1.8) & 30.04 (0.45)  &  \\
\hline
\end{tabular}
\label{tab1}
\end{table*}

The data were converted from the native $lta$ format into $FITS$ format and imported into $AIPS$ 
\footnote{AIPS is produced and maintained by the National Radio Astronomy Observatory, a facility of the 
National Science Foundation operated under cooperative agreement by Associated Universities, Inc.} 
These data were calibrated and imaged using 25 facets at 610 MHz and 49 
facets at 235 MHz.  The synthesized beamshapes were elliptical with typical sizes 
being $10'' \times5''$ at 610 MHz and $20''\times10''$ at 235 MHz with a position angle $\sim 30^{\circ}$.

\begin{figure}
\includegraphics[width=6.0cm]{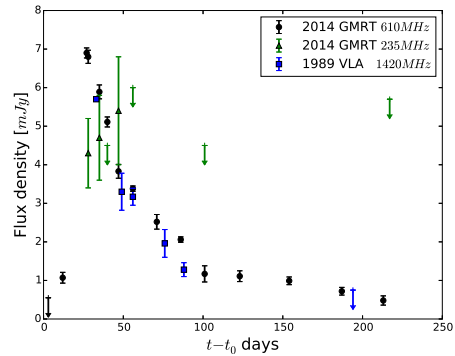}
\caption{ 
Light curve of V745 Scorpii: 610  MHz and
235 MHz from the 2014 outburst using GMRT and 1.4 GHz from the previous outburst in 1989 using
VLA archival data.   t$_0$ is 6 February 2014.  }
\label{fig1}
\end{figure}

The results of GMRT observations of V745 Sco are listed in Table \ref{tab1} and shown in Fig. \ref{fig1}.  
The first detection was on day 12 (18/2/2014) after the recorded outburst.
Due to a position offset in the map, \citet{2014ATel.5962....1K} missed reporting the detection on day 12. 
Observations on 6 March 2014 detected a 
strong radio source at both 610 MHz and 235 MHz (Table \ref{tab1}).  No correlation between the varying 
strength of the target source and the calibrators is observed (Table \ref{tab1}).

We analysed the VLA archival data of V745 Sco from 1989 and the results are  
shown in Table \ref{tab2} and Fig. \ref{fig1}.  The data on RS Ophiuchi are obtained from 
literature -- the 1.4 GHz data from the 1985 outburst is from \citet{1986ApJ...305L..71H} and the model 
fit to the 2006 outburst is from \citet{2007ApJ...667L.171K}.

\section{Interpreting the observations and modelling the light curves} V745 Sco is the second symbiotic 
recurrent nova which has been studied at GMRT frequencies where the emission is predominantly 
synchrotron in nature. The peak radio power estimated for a distance of $7.8\pm1.8$ kpc 
\citep{2010ApJS..187..275S} at 610 and 235 MHz are respectively $4.7\times10^{13}$ W Hz$^{-1}$ and 
$4\times10^{13}$ W Hz$^{-1}$. For comparison, the peak power from RS Ophiuchi at 610 MHz following its 
outburst in 2006 \citep{2007ApJ...667L.171K} was $1.5\times10^{13}$ W Hz$^{-1}$.

Assuming equipartition of energy between the relativistic particles and magnetic field, expected under 
minimum energy condition, we estimate a magnetic field of 0.03 G and energies of $\sim10^{32}$ Joule 
for the 2014 outburst of V745 Sco. An emitting shell of radial extent 1 AU was assumed at a distance of 30 
AU from the white dwarf as inferred from the near-infrared \citep{2014ApJ...785L..11B} and X-ray 
observations \citep{2015MNRAS.448L..35O}. A magnetic field of strength 0.04 G with 
energy in relativisitic particles being $2.8\times10^{31}$ Joule ie about 0.02\% of total energy was 
estimated for the 1985 outburst in RS Ophiuchi \citep{1985MNRAS.217..205B}.  The magnetic field strengths 
and particle energies match to within an order of magnitude in the two recurrent novae and for two 
consecutive outbursts.

\begin{figure*}
\includegraphics[width=6.0cm]{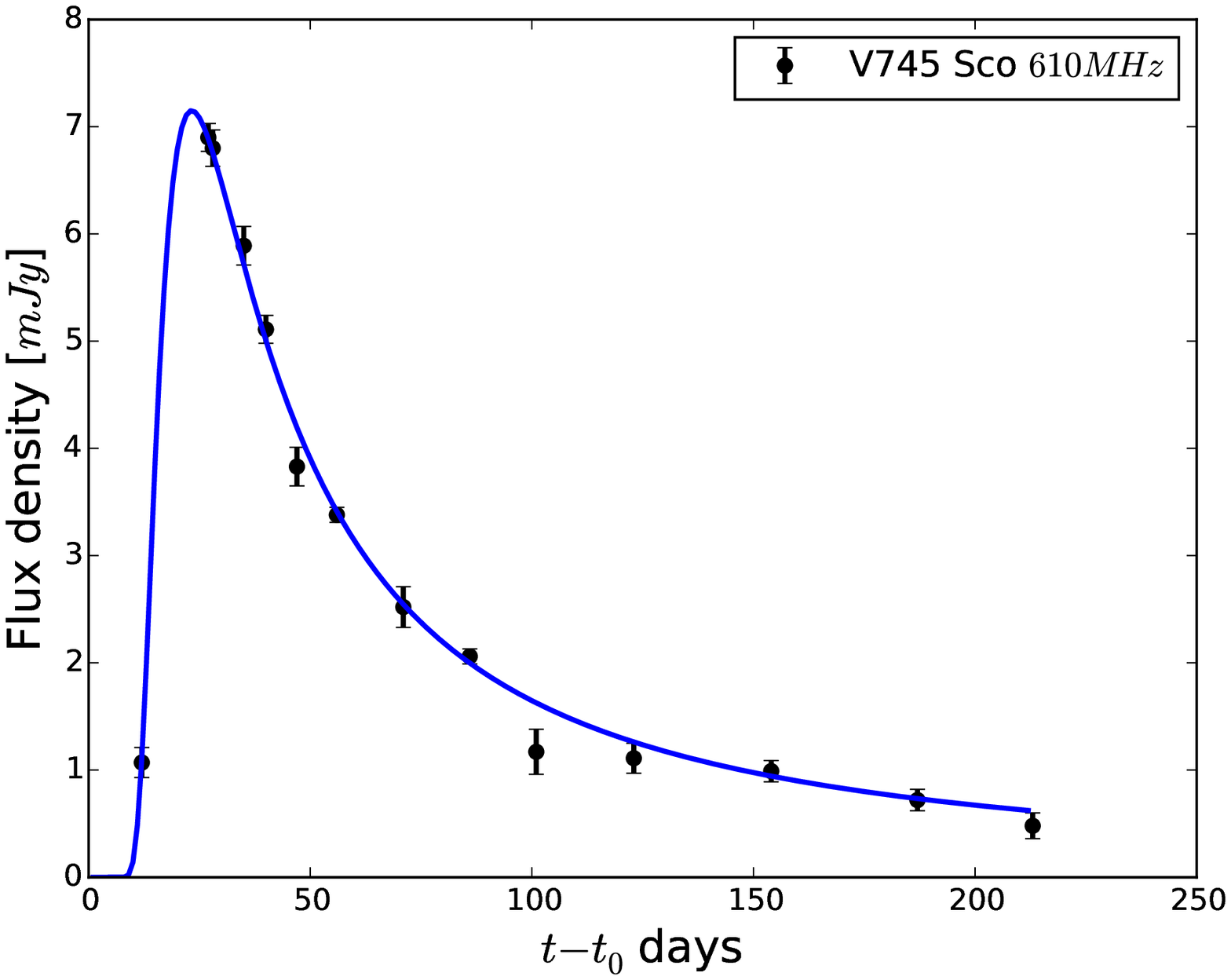}
\includegraphics[width=6.0cm]{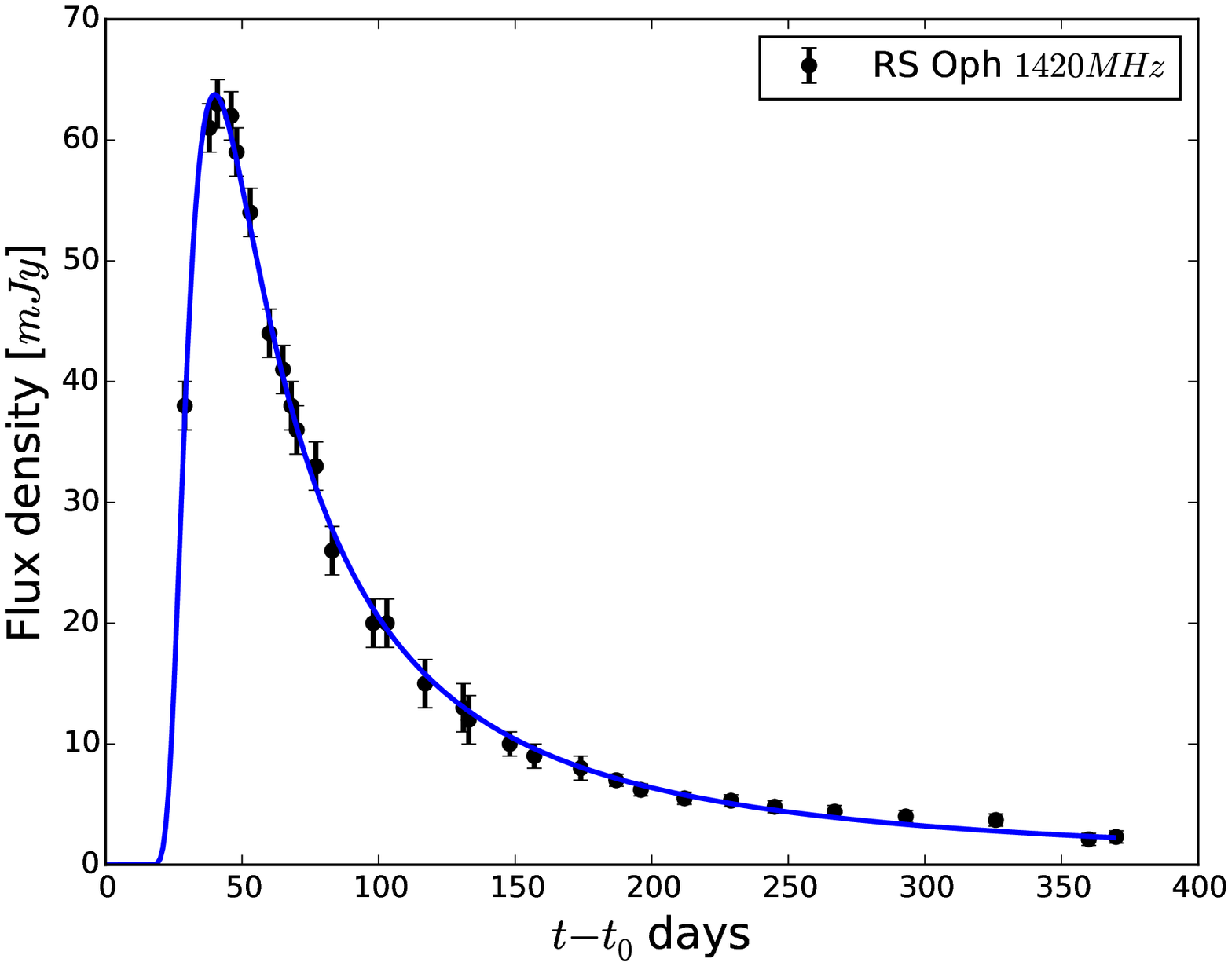}
\caption{(left) The best model fit (solid line) to the light curve of V745 Sco at 610 MHz from
its outburst in 2014. 
(right) The best model fit (solid line) to the light curve of RS Ophiuchi at 1.4 GHz from its
outburst in 1985.  Data points are taken from \citet{1986ApJ...305L..71H}. }
\label{fig2}
\end{figure*}

We use the parameteric model that has been presented in Eqn. 1 in \citet{2002ARA&A..40..387W} for 
explaining supernova light curves.  The assumptions in the model are implicitly included and we do 
not explore any possible differences due to the observed bipolar nature of 
synchrotron emission from RS Ophiuchi in its 1985 outburst \citep{1989MNRAS.237...81T} and its 2006 
outburst \citep{2006Natur.442..279O}.  In the \citet{2002ARA&A..40..387W} model, there is a 
frequency-dependent delay in the detection of synchrotron emission due to the opacity of the foreground 
thermal gas.  The radio emission rises as opacity decreases with peak emission at unity opacity and 
then declines as the emitting shell expands. The model includes 
opacity due to several different components.  We only included the local opacities due to the uniform and 
clumpy parts of the circumbinary material which well explained the light curves from the 2006 
outburst in RS Ophiuchi \citep{2007ApJ...667L.171K}.  This model was fitted to the light curves at 610 
MHz from the 2014 outburst (Fig. \ref{fig2}) and at 1.4 GHz from the 1989 outburst in V745 Sco and
to the 1.4 GHz data from the 1985 outburst in RS Ophiuchi (Fig. \ref{fig2}).  
The model outputs are listed in Table \ref{tab3}.  

\subsection{The free-free optical depth variation in V745 Sco } 

\begin{figure}
\includegraphics[width=6.0cm]{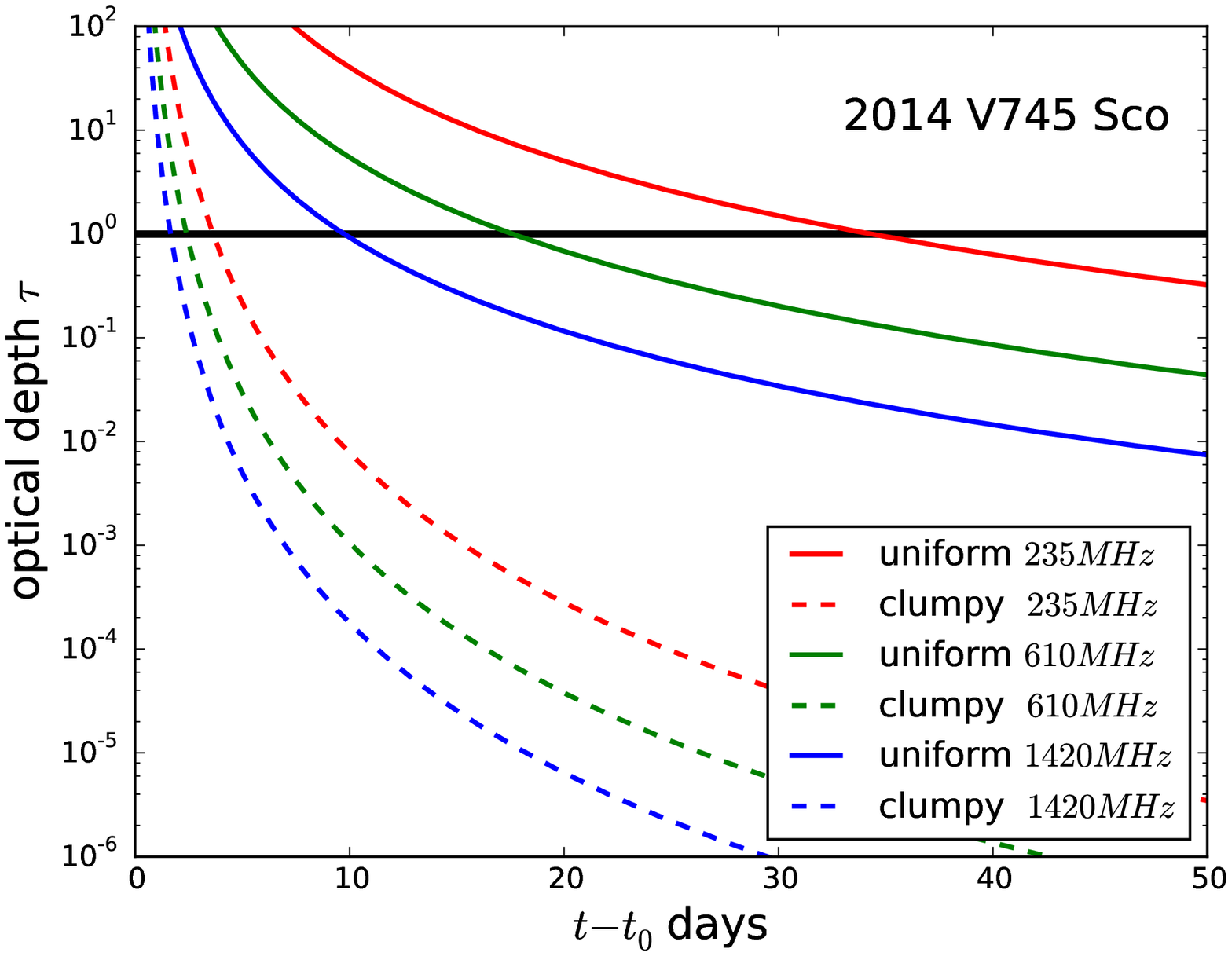}
\includegraphics[width=6.0cm]{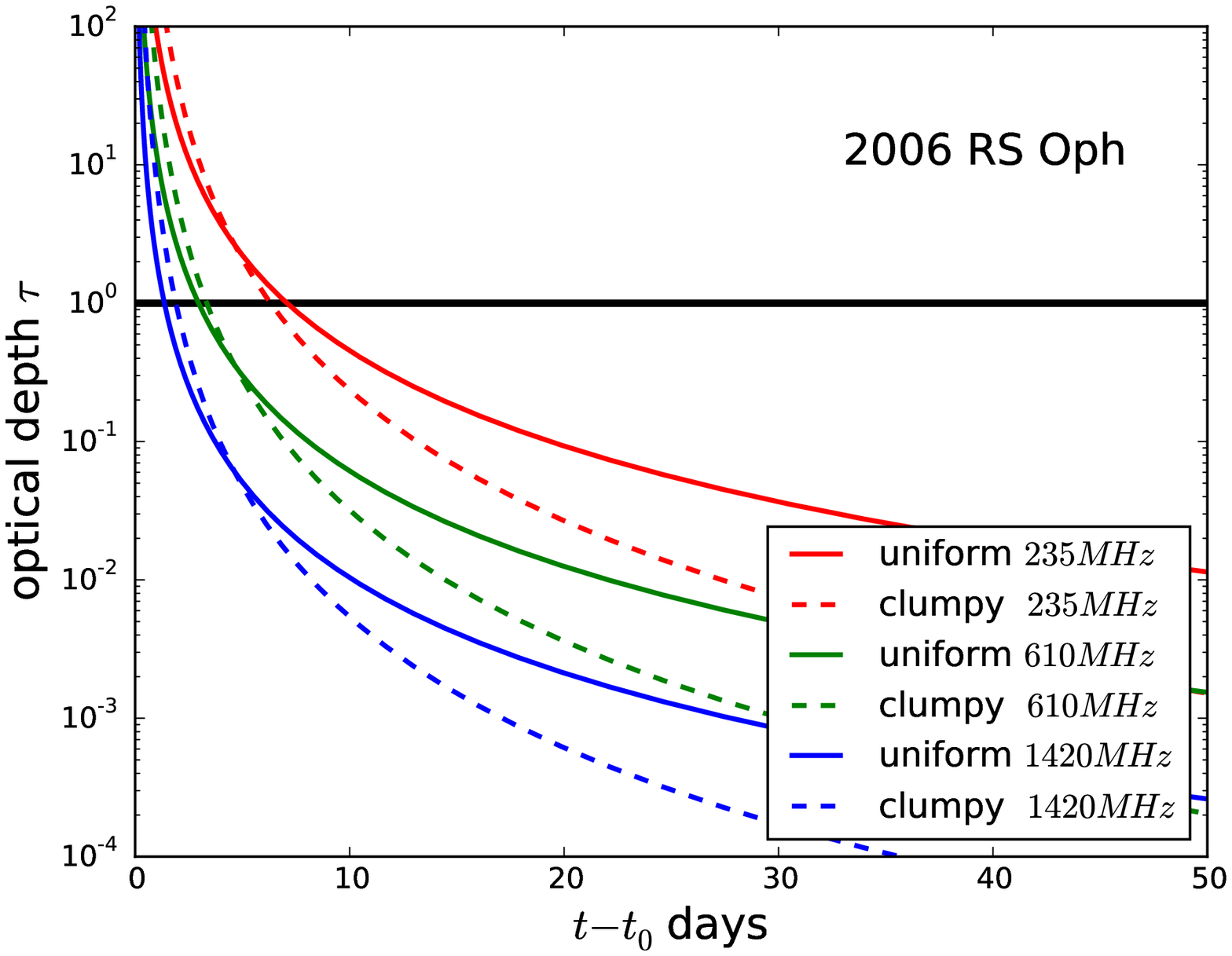}
\includegraphics[width=6.0cm]{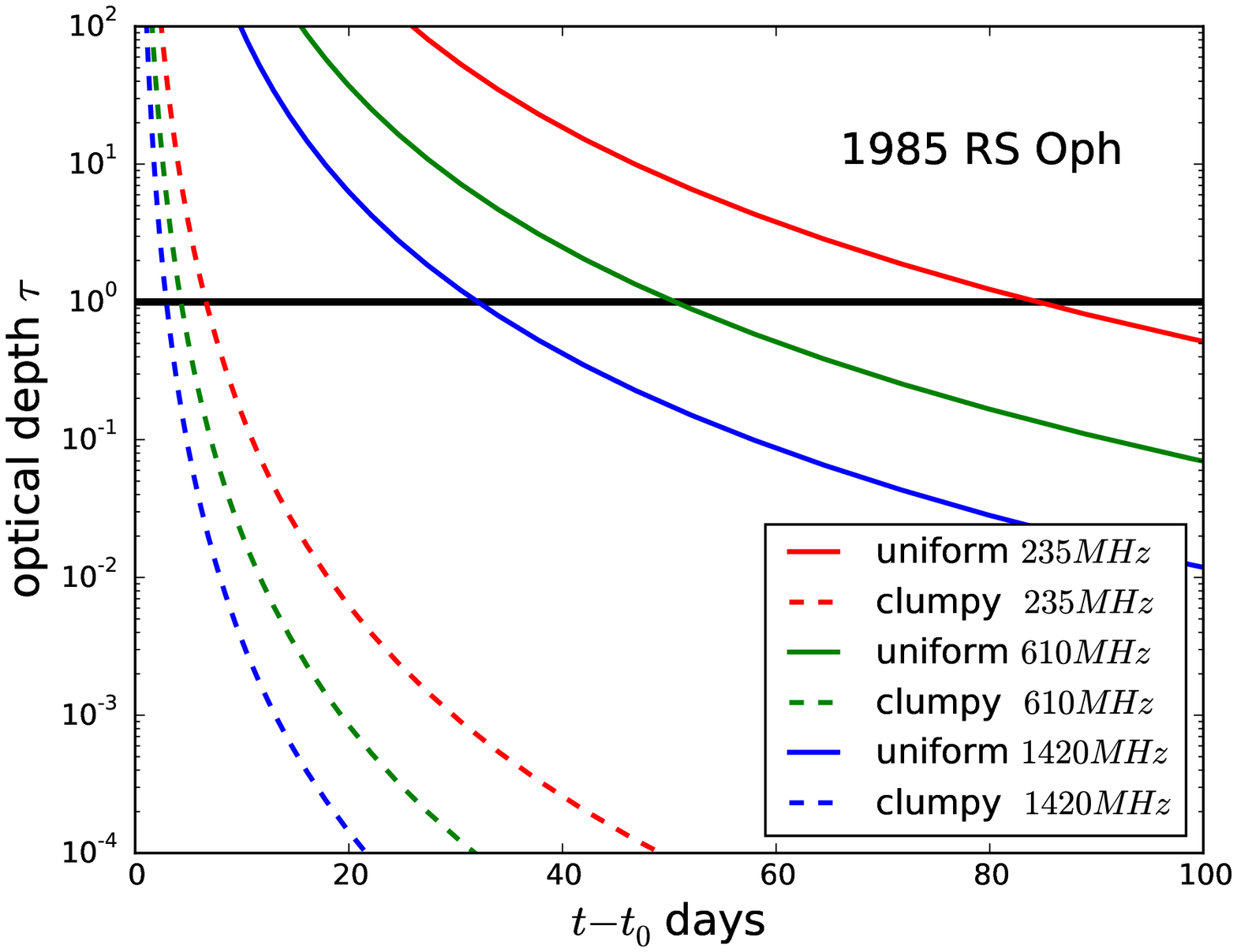}
\caption{ The evolution of the optical depth due to the uniform density and clumpy medium in 
the nova system for (top) 2014 outburst of V745 Sco,
(centre) 2006 outburst of RS Oph using parameters given in \citet{2007ApJ...667L.171K},
(bottom) 1985 outburst of RS Ophiuchi.   Clumpy medium shows similar behaviour for the outbursts
in 1985 and 2006 whereas the uniform density medium shows evolution.  The large optical depths
are not physical and are shown only to indicate total absorption.   The region of interest is
close to where opacity is one as indicated by the black horizontal line. 
  } 
\label{fig3}
\end{figure}

Using the model parameters listed in Table \ref{tab3}, 
we have plotted the variation in the optical depth with time (see Eqn 1 and Fig \ref{fig3} top). 
The temporal variation in the optical depth due to uniformly distributed gas,
$\rm \tau_{uniform}$ and the clumpy gas, $\rm \tau_{clumpy}$ is \citep{2002ARA&A..40..387W}: 

\begin{equation}
\rm
\tau_{uniform}[\tau_{clumpy}] = K_2 [K_3]\left(\frac{\nu}{5GHz}\right)^{-2.1}\left(\frac{t-t_0}{1day}\right)^{\delta_{uniform}[\delta_{clumpy}]} 
\end{equation}

\noindent 
K$_2$, K$_3$, $\delta$ are determined from the 
fits to the light curve (see Table \ref{tab3}). The variation in both optical depths at three frequencies 
235 MHz, 610 MHz and 1.4 GHz is shown in Fig. \ref{fig3}.  As seen in the figure, $\rm \tau_{clumpy}$ 
drops to one within five days following the outburst at all bands. $\rm \tau_{uniform}$ shows a slower 
decline -- is one around day 10 at 1.4 GHz, day 18 at 610 MHz and day 35 at 235 MHz which would 
roughly correspond to the peak emission at those bands. Our observations at 235 MHz are consistent with 
this model.  Thus the turnon of the synchrotron emission at the low radio frequencies is determined 
primarily by the optical depth of the uniform density gas in the 2014 outburst of V745 Sco.  Due to the absence
of data points leading to the peak of the 1989 outburst (see Table \ref{tab3}), all the model parameters are not 
well constrained and do not allow a study of the opacity variation.  The temporal flux variation is 
similar in both epochs.  The model fit predicts the peak at 1.4 GHz in 1989 to have been around day 18 as 
compared to day 10 in 2014.  This suggests evolution in the environment of the nova.

\begin{table}
  \caption{Analysis of VLA archival data (AH 383, AH 389) at 1.4 GHz on V745 Sco following the outburst
in 1989.  t$_0$ is 30 July 1989.  }
  \begin{tabular}{@{}llll@{}}
  \hline
t-t$_0$  &  date  & S  & Reference   \\
days  &         &  mJy &  \\
\hline
33 & 01 Sep 1989 &    5.7       &  \citet{1989IAUC.4853....2H}   \\
49 & 17 Sep 1989  &   3.3(0.48)    &  VLA archives  \\
56 & 24 Sep 1989   &  3.17(0.22)    &  VLA archives  \\
76 & 07 Oct 1989    & 1.96(0.36)    &  VLA archives   \\
88 & 19 Oct 1989    & 1.28(0.18)    &  VLA archives  \\
194 & 02 Feb 1990     & $<0.75$       &  VLA archives  ($3\sigma$)  \\
\hline
\end{tabular}
\label{tab2}
\end{table}

\begin{table*}
\caption{Model outputs. The spectrum is assumed to have a spectral index of $\alpha=-0.5$.  Emission measure
of absorbing gas when the optical depth is one.
The electron temperature is assumed as $10^4$ K. N is the number of detections,
$\rm t_{peak}$,  $\rm S_{peak}$,  $\rm t_{1\%}$ refer to the day on which peak emission occurred,
peak emission strength and the day on which the emission was $1\%$ of peak emission. $\beta$ is the temporal decay index,
K$_2$, $\rm \delta_{uniform}$, K$_3$, $\rm \delta_{clumpy}$ are the constants from Eq. 1.  }
\begin{tabular}{l|c|c|c|c|c|c|c|c|c|c|c|c|c}
\hline
Nova & Outburst & $\nu$ &N &   $\rm t_{peak}$ & $\rm S_{peak}$ & $\rm t_{1\%}$ & $\rm \beta$ & K$\rm_2$ & $\rm \delta_{uniform}$ & K$\rm_3$ & $\rm \delta_{clumpy}$  & $\rm \chi^2_{red}$ & EM$\rm _{peak}$ \\
      & epoch    &     &   & days  & mJy & days & & & & & &  &cm$\rm ^{-6} $pc \\
\hline
V745 Sco & 2014  & 610 MHz & 14 & 23 &  7.1 & 9.5 & $-1.3$ & 65.8 & $-3$ & $0.8$ & $-4.8$ & 1.2 & $10^6$  \\
         & 1989  & 1.4 GHz & 5 & 18 & 8.7 & 7 & $-1.4$ & - & $-2.9$ & - & $-6.3$ & 4 & $6\times10^6$  \\
RS Ophiuchi & 2006$^1$ & 610 MHz & 14 & 29 & 52.9 & 6 & $-1.24$ & 0.14 & $-2.29$ & 0.53 & $-3.14$ & 1.5 &  $10^6$ \\
            & 1985$^2$ & 1.4 GHz & 30 & 40 & 63.8 & 20.5 & $-1.7$ & 53060 & $-3.9$ & 9.8 & $-4.6$ &  0.54 & $6\times10^6$  \\
\hline
\end{tabular}

{\small $^1$ Parameters from the multi-frequency light curve fits in \citet{2007ApJ...667L.171K}.}
{\small $^2$ The 1985 light curve data at 1.4 GHz is taken from \citet{1986ApJ...305L..71H}}
\label{tab3}
\end{table*}

\subsection{The free-free optical depth variation in RS Ophiuchi} We also did a similar study of opacity 
variation in RS Ophiuchi using our model fit to the 1.4 GHz data from the 1985 outburst taken from 
\citet{1986ApJ...305L..71H} and the \citet{2007ApJ...667L.171K} fit parameters for the 2006 outburst 
(Fig. \ref{fig3}). Interestingly, 
the variation in opacities due to uniform density gas and clumpy gas is comparable following the outburst 
in 2006 whereas the 1985 fit shows a rapid fall in $\rm \tau_{clumpy}$ as noted for V745 Sco in 2014. The 
differences in the 1985 and 2006 outbursts of RS Ophiuchi are (1) The turn-on in 2006 at a given frequency 
occurs at an earlier date following the outburst as compared to the 1985 outburst.  (2) The turn-on day in 
1985 is primarily determined by the uniform density gas whereas in 2006, it is determined by both the 
uniform density and clumpy gas.  (3) The $\tau_{uniform}$ in 1985 falls to unity around day 32 at 1.4 GHz and 
day 85 at 235 MHz whereas in 2006, it is unity around day 2 at 1.4 GHz and around day 8 at 
235 MHz.  We infer the net effect to be reduction in the ambient ionized gas densities in 2006 as 
compared to 1985. \citet{2007ApJ...667L.171K} had arrived at a similar result  
using the 325 MHz data from 2006 when it was detected on day 38 and using the result from 1985 
that no emission was detected upto day 66 \citep{1987MNRAS.224..791S}  and inferred that the 
absorbing densities in 2006 were about $30 \%$ of those in 1985.  The model fitted to the 1985 1.4 GHz 
data on RS Ophiuchi is consistent with this conclusion and suggests that the emission at 325 MHz would 
have been visible $\sim$ day 66 in 1985.

\subsection{Comments on evolution of novae from opacity variation} Our main results from the synchrotron 
light curve fitting for both the novae are: (1) earlier turn-on and peaks observed at a given frequency 
following an outburst for successive outbursts, (2) clumpy material distributed closer to the system
and uniformly distributed gas being more widespread. In Table \ref{tab3}, we have listed the 
emission measure (EM) of the foreground thermal gas when it would be 
transparent to the emission at 610 MHz and 1.4 GHz.  The earlier peaks with successive outbursts would 
indicate reducing EM of the absorbing gas.

Theoretical studies indicate that when the accretion rate exceeds some critical limit (few times $ 
10^{-7}$ $\rm~ M_\odot yr^{-1}$; \citealt{2001ApJ...558..323H}), then the envelope on the white dwarf can 
expand to the size of a red giant \citep{1982ApJ...253..798N}.  This can then lead to the formation of a 
common envelope (e.g. \citealt{1979PASJ...31..287N}) which can trigger a spiral-in of the binary and a 
double degenerate system \citep{1984ApJ...284..719I}.  \citet{1996ApJ...470L..97H} found that another 
outcome of the larger accretion rate would be fast optically thick winds ($\rm \sim 1000 kms^{-1}$, $\ge 
10^{-6} \rm~ M_\odot yr^{-1}$) blown by the white dwarf which they refer to as accretion winds.  The 
accretion winds would stabilise the system and the binary can continue to evolve as a single degenerate 
system \citep{1996ApJ...470L..97H}.  The accretion rates for V745 Sco and RS Ophiuchi were estimated to be 
$\rm 2\times10^{-7}~ M_\odot~yr^{-1} $ (using a recurrence timescale of 25 years) and $\rm 1.2 \times 
10^{-7}~ M_\odot~yr^{-1}$ \citep{2001ApJ...558..323H}.  Combining our results with the theoretical 
arguments, we infer the following: \\ 1. The onset of 
synchrotron radio emission is delayed by the optically thick winds blown by the white dwarf which constitutes 
the uniform density component.  The clumpy component rapidly gets transparent.  We suggest this to be due 
to the material from the accretion disk close to the white dwarf which is blown off in each outburst 
\citep{2001ApJ...558..323H}. \\ 2. The earlier turnon of the radio emission with successive outbursts 
would then indicate the reduction in emission measure of the accretion winds.  This could imply a reduced 
accretion rate ($\rm < ~few~ times~ 10^{-7}$ $\rm~ M_\odot yr^{-1}$) on the white dwarf.  The estimated 
accretion rate on both the novae is about $10^{-7}$ $\rm~ M_\odot yr^{-1}$. If indeed the accretion rate 
has dropped causing the winds to stop -- either the two stars can spiral-in leading to a double degenerate 
system or if the white dwarf is massive enough, it can explode as a type 1a supernova.  If the latter is 
not the case, then there are reasons to believe that the single degenerate system might not evolve into a 
type 1a supernova.  \\ 3.  Alternately since the mass of the white dwarfs in both systems is believed 
to be close to the Chandrasekhar limit, the critical accretion rate limit is larger at $10^{-6} \rm~ 
M_\odot~yr^{-1}$ \citep{1982ApJ...253..798N}.  Since the estimated accretion rates are lower than this 
larger critical limit, it would cause the winds to stop as the white dwarf grows in mass, leading to a 
transparent ambience at radio wavelengths.  From 
our results on V745 Sco and RS Oph over two outbursts, it can be surmised that the synchrotron emission at 
610 MHz in the next outburst from V745 Sco should be detected before day 9.5 and from RS Oph before day 6.  
If the accretion rate has reduced, then it would lengthen the period between two outbursts -- however if 
only the accretion winds have stopped, this should have no effect on the outburst frequency.  \\ 4. The 
electron energy spectrum set up by the shock in two distinct outbursts appear similar for the two systems 
studied here.  Multifrequency radio synchrotron data is required for further study which is feasible 
in future outbursts in the fast-evolving recurrent nova systems provided time allocation is made faster 
on major radio telescopes. 

\section{Summary and conclusion}
In this paper we have presented our observations, at 610 and 235 MHz using the GMRT, of the recurrent nova 
system V745 Sco following its outburst in 2014. 
The parametric model including opacities due to clumpy and uniform media in \citet{2002ARA&A..40..387W} 
explains the light curves of V745 Sco and RS Ophiuchi. 
We conclude the following from our study:

\noindent
(1) The radio synchrotron emission is visible sooner after the outburst, with each outburst. In V745 Sco, 
the 610 MHz emission peaked $\sim$ day 23 in 2014 and $\sim$ day 18 at 1.4 GHz in the 1989 outburst. Our 
model fit predicts that the 1.4 GHz emission would have peaked $\sim$ day 10 in 2014. In RS Ophiuchi, the 
radio synchrotron emission at 1.4 GHz turned on on day 20.5 in 1985 whereas the first detection in 2006 
was on day 4.7 \citep{2009MNRAS.395.1533E}.

\noindent
(2) The circumbinary material in the recurrent nova with a red giant companion is evolving with time. 
Clumpy material lies closer to the system compared to the extent of the uniform medium.
This material could be due to the accretion disk of 
the white dwarf which is destroyed with each outburst.  The uniform density component is caused by the hot 
optically thick winds blowing from the white dwarf.  The earlier visibility could indicate that the winds 
are arrested due to the accretion rate falling below some critical rate for a given white dwarf mass.  
This could lead to multiple evolutionary scenarios which need to be investigated further. 
Interestingly, \citet{2013AJ....146...55W} also required a medium with clumpy and uniform components to
explain optical and X-ray data. Well-sampled multifrequency data
during the rise of the light curve to peak are necessary to estimate
the effect of the uniform and clumpy components.

\noindent
(3) All recurrent nova systems at all wavebands in quiescent 
(e.g. \citealt{1999A&A...344..177A}) and outburst phases need to be studied.  
Novae are an important Galactic system suited 
to the study of shock interaction with the ambient medium and its evolution
over short timescales and multiple epochs. 

\noindent
\section*{Acknowledgements}
We thank the reviewer, A.~R.~Taylor for a helpful review. We thank the staff of the GMRT that made these 
observations possible. GMRT is run by NCRA of the Tata Institute of Fundamental Research. We thank the 
Centre Director, NCRA for granting DDT time.  We thank the AAVSO for all their valuable work.  NGK thanks 
Prasad Subramanian for discussions and Dave Green for comments on the manuscript. PD acknowledges that 
this work is partially supported by the DST INSPIRE Faculty Fellowship award [IFA-13 PH 54] and performed 
at IISER, Bhopal.


\begin{thebibliography}{99}

\bibitem[\protect\citeauthoryear{Anupama 
\& Miko{\l}ajewska}{1999}]{1999A&A...344..177A} Anupama G.~C., Miko{\l}ajewska J., 1999, A\&A, 344, 177

\bibitem[\protect\citeauthoryear{Banerjee et 
al.}{2014}]{2014ApJ...785L..11B} Banerjee D.~P.~K., Joshi V., Venkataraman 
V., Ashok N.~M., Marion G.~H., Hsiao E.~Y., Raj A., 2014, ApJ, 785, L11 

\bibitem[\protect\citeauthoryear{Bode 
\& Kahn}{1985}]{1985MNRAS.217..205B} Bode M.~F., Kahn F.~D., 1985, MNRAS, 217, 205 

\bibitem[\protect\citeauthoryear{Cheung, Jean, 
\& Shore}{2014}]{2014ATel.5879....1C} Cheung C.~C., Jean P., Shore S.~N., 2014, ATel, 5879

\bibitem[\protect\citeauthoryear{Duerbeck}{1989}]{1989Msngr..58...34D} 
Duerbeck H.~W., 1989, Msngr, 58, 34

\bibitem[\protect\citeauthoryear{Eyres et al.}{2009}]{2009MNRAS.395.1533E} 
Eyres S.~P.~S., et al., 2009, MNRAS, 395, 1533 

\bibitem[\protect\citeauthoryear{Hachisu, Kato, 
\& Nomoto}{1996}]{1996ApJ...470L..97H} Hachisu I., Kato M., Nomoto K., 1996, ApJ, 470, L97 

\bibitem[\protect\citeauthoryear{Hachisu 
\& Kato}{2001}]{2001ApJ...558..323H} Hachisu I., Kato M., 2001, ApJ, 558, 323 

\bibitem[\protect\citeauthoryear{Hjellming et 
al.}{1986}]{1986ApJ...305L..71H} Hjellming R.~M., van Gorkom J.~H., Taylor 
A.~R., Sequist E.~R., Padin S., Davis R.~J., Bode M.~F., 1986, ApJ, 305, L71

\bibitem[\protect\citeauthoryear{Hjellming}{1989}]{1989IAUC.4853....2H} 
Hjellming R.~M., 1989, IAUC, 4853, 2 

\bibitem[\protect\citeauthoryear{Iben 
\& Tutukov}{1984}]{1984ApJ...284..719I} Iben I., Jr., Tutukov A.~V., 1984, ApJ, 284, 719 

\bibitem[\protect\citeauthoryear{Kantharia et 
al.}{2014}]{2014ATel.5962....1K} Kantharia N.~G., Roy N., Anupama G.~C., 
Banerjee D.~P.~K., Ashok N.~M., Dutta P., Prabhu T.~P., Johri A., 2014, 
ATel 5962  

\bibitem[\protect\citeauthoryear{Kantharia et 
al.}{2007}]{2007ApJ...667L.171K} Kantharia N.~G., Anupama G.~C., Prabhu 
T.~P., Ramya S., Bode M.~F., Eyres S.~P.~S., O'Brien T.~J., 2007, ApJ, 667, 
L171 

\bibitem[\protect\citeauthoryear{Nomoto, Nariai, 
\& Sugimoto}{1979}]{1979PASJ...31..287N} Nomoto K., Nariai K., Sugimoto D., 1979, PASJ, 31, 287 

\bibitem[\protect\citeauthoryear{Nomoto}{1982}]{1982ApJ...253..798N} Nomoto 
K., 1982, ApJ, 253, 798 

\bibitem[\protect\citeauthoryear{O'Brien et 
al.}{2006}]{2006Natur.442..279O} O'Brien T.~J., et al., 2006, Natur, 442, 
279 

\bibitem[\protect\citeauthoryear{Orio et al.}{2015}]{2015MNRAS.448L..35O} 
Orio M., Rana V., Page K.~L., Sokoloski J., Harrison F., 2015, MNRAS, 448, 
L35 

\bibitem[\protect\citeauthoryear{Schaefer}{2010}]{2010ApJS..187..275S} 
Schaefer B.~E., 2010, ApJS, 187, 275 


\bibitem[\protect\citeauthoryear{Spoelstra et 
al.}{1987}]{1987MNRAS.224..791S} Spoelstra T.~A.~T., Taylor A.~R., Pooley 
G.~G., Evans A., Albinson J.~S., 1987, MNRAS, 224, 791 

\bibitem[\protect\citeauthoryear{Swarup et al.}{1991}]{1991CuSc...60...95S} 
Swarup G., Ananthakrishnan S., Kapahi V.~K., Rao A.~P., Subrahmanya C.~R., 
Kulkarni V.~K., 1991, CuSc, 60, 95

\bibitem[\protect\citeauthoryear{Taylor et al.}{1989}]{1989MNRAS.237...81T} 
Taylor A.~R., Davis R.~J., Porcas R.~W., Bode M.~F., 1989, MNRAS, 237, 81 

\bibitem[\protect\citeauthoryear{Weiler et 
al.}{2002}]{2002ARA&A..40..387W} Weiler K.~W., Panagia N., Montes M.~J., Sramek R.~A., 2002, ARA\&A, 40, 387 

\bibitem[\protect\citeauthoryear{Williams}{2013}]{2013AJ....146...55W} 
Williams R., 2013, AJ, 146, 55 



\end{thebibliography}
\end{document}